\RequirePackage{fix-cm}
\documentclass[smallextended, natbib]{svjour3}       
\setcitestyle{square}
\smartqed  
\usepackage{graphicx}

\usepackage{booktabs}
\usepackage{subcaption}
\newcommand{\ra}[1]{\renewcommand{\arraystretch}{#1}}
\usepackage{etoolbox, siunitx}
\renewcommand{\bfseries}{\fontseries{b}\selectfont}
\newrobustcmd{\B}{\bfseries}
\usepackage{amsmath,hyperref}
\usepackage{comment}
\newenvironment{conditions}
  {\par\vspace{\abovedisplayskip}\noindent\begin{tabular}{>{$}l<{$} @{${}={}$} l}}
  {\end{tabular}\par\vspace{\belowdisplayskip}}

\newenvironment{conditions2}
  {\par\vspace{\abovedisplayskip}\noindent\begin{tabular}{>{$}l<{$} @{${}{}$} l}}
  {\end{tabular}\par\vspace{\belowdisplayskip}}
\begin{document}

\title{Deep learning approach to left ventricular non-compaction measurement
}


\author{Jes\'us M. Rodríguez-de-Vera    \and Josefa Gonz\'alez-Carrillo \and Jos\'e M. Garc\'ia \and Gregorio Bernab\'e  
}


\institute{Jes\'us M. Rodríguez-de-Vera and Jos\'e M. Garc\'ia and Gregorio Bernab\'e  \at
              Computer Engineering Department. University of Murcia, Murcia (Spain) \\
              \email{jesus.molinar@um.es, jmgarcia@um.es, gbernabe@um.es}           
           \and
           Josefa Gonz\'alez-Carrillo \at
              Hospital Virgen de la Arrixaca, Murcia (Spain)\\
              \email{josegonca.alarcon@gmail.com}
}

\date{Received: date / Accepted: date}

\maketitle

\begin{abstract}
Left ventricular non-compaction (LVNC) is a rare cardiomyopathy characterized by abnormal trabeculations in the left ventricle cavity. Although traditional computer vision approaches exist for LVNC diagnosis, deep learning-based tools could not be found in the literature. In this paper, a first approach using convolutional neural networks (CNNs) is presented. Four CNNs are trained to automatically segment the compacted and trabecular areas of the left ventricle for a population of patients diagnosed with Hypertrophic cardiomyopathy. Inference results confirm that deep learning-based approaches can achieve excellent results in the diagnosis and measurement of LVNC. The two best CNNs (U-Net and Efficient U-Net B1) perform image segmentation in less than \SI{0.2}{\second} on a CPU and in less than \SI{0.01}{\second} on a GPU. Additionally, a subjective evaluation of the output images with the identified zones is performed by expert cardiologists, with a perfect visual agreement for all the slices, outperforming already existing automatic tools.
\end{abstract}

\section{Introduction}
\label{intro}
Left ventricular non-compaction (LVNC) is a recently classified form of cardiomyopathy, characterized by abnormal trabeculations in the left ventricle cavity \cite{towbin2015lvnc}. It is a rare condition associated with different types of heart defects: in \cite{STAHLI20132477}, 12\% of patients with LVNC had other congenital cardiac diseases. In \cite{associatedLVNC}, 93\% of patients diagnosed with ventricular non-compaction had another cardiomyopathy.
 
There is still no consensus in the medical community on the way to quantify and value LVNC \cite{towbin2015lvnc}. Some of the already existing approaches are the measurement of the trabecular mass based on endocardium delineation described in \cite{jacquier2010measurement}, or the estimation of endocardial borders complexity using fractal methods proposed in \cite{captur2014abnormal, captur2013quantification}. In \cite{gbernabe, BERNABE2015610}, an automatic software tool based on medical experience was developed to quantify the degree of non-compaction using cardiac MRI. This tool uses traditional computer vision methods and requires manual tuning by the cardiologists. In \cite{gbernabe}, a self-optimization software tool was proposed, achieving no diagnostically significant differences in 81.34\% of the images evaluated.

The irruption of deep learning (DL) techniques in the last years has led to great improvements in medical image analysis, surpassing existing state-of-the-art approaches \cite{litjens2017survey}. Particularly, cardiac MRI segmentation using DL has attracted considerable interest, with a growing number of publications on the subject \cite{cardiacReview}. Nevertheless, existing papers focus mostly on finding main heart structures such as ventricles and atria (where DL methods have equalled and even surpassed expert analysis \cite{bernard:hal-01803621}). There are fewer works on segmenting abnormal cardiac tissue regions which are related to some 
heart diseases, such as trabeculae and LVNC.

In this work, we present a DL-based approach to the measurement and diagnosis of LVNC. This project aims to take advantage of the significant progress in medical image segmentation to the estimation and diagnosis of LVNC. To this end, four different convolutional neural networks (CNNs) have been trained, and they have been evaluated in terms of its segmentation accuracy and the usefulness of their output to cardiologists. To the best of our knowledge, this is the first attempt to apply DL to the problem of measuring hyper-trabeculation levels, so we present this paper as a possible starting point for an accurate, reliable, and fully automated diagnosis of this cardiomyopathy.

Section \ref{sec:background} presents different deep learning techniques already used for image segmentation generally, and particularly for cardiac segmentation. Section \ref{sec:methods} describes data preprocessing, and the neural network architectures and training methods applied to this project. The obtained performance of the proposed method, along with its segmentation and diagnosis ability, can be found in Section \ref{sec:results}, where our solution is also compared with already existing techniques. Finally, Section \ref{sec:conclusion} summarizes the work, presents the main conclusions of the paper, and proposes a future path of research.

\section{Background}\label{sec:background}

One of the most critical elements of any training process is the data fed to the model. In particular, most of the current challenges of medical image analysis are related to data itself \cite{cardiacReview, Hesamian2019, litjens2017survey}. Therefore, the first element of the image processing pipeline is usually focused on overcoming these recurrent problems. When talking about medical image segmentation, the scarcity of annotated images makes it difficult to create models with generalization capability, and thus, overfitting occurs \cite{cardiacReview, Hesamian2019}. Multiple approaches to mitigate this difficulty are found in the literature, being pre-processing and data augmentation the most prominent ones \cite{litjens2017survey}.

Given that the most usual approach to cardiac image segmentation is semantic segmentation, the most widely used loss functions treat the problem as a pixel-wise classification task like cross-entropy loss. For example, it is usual to use loss functions that are intended to optimize the Dice coefficient, like the generalized Dice loss \cite{Sudre_2017} and the Lovász-Softmax loss, presented in \cite{berman_triki_blaschko_2018}. Besides being a differentiable surrogate function of Jaccard index, minimization of Lovász-Softmax loss also leads to Dice coefficients optimization, achieving better results than Dice loss in some medical image segmentation tasks \cite{10.1007/978-3-030-32245-8_11}.

On the other hand,  there also exist loss functions based on borders differences instead of regions \cite{kervadec2018boundary} (boundary losses), which make network output to be more anatomically suitable \cite{kervadec2018boundary}. That is important because even state-of-the-art cardiac segmentation solutions produce anatomically impossible results. This problem is more frequent in apical and basal slices than in mid-cavity images \cite{bernard:hal-01803621}. Moreover, the boundary losses also help to prevent the class imbalance problem, which is particularly critical since that small tissues are usually essential for the diagnosis.

CNNs constitute state-of-the-art for medical image segmentation (and cardiac MRI segmentation in particular). One of the first architectures specifically designed for medical image segmentation is the well-known U-Net \cite{ronneberger_fischer_brox_2015}. This network has an encoder-decoder structure, in which skip connections between encoder and decoder are used to recover spatial context loss and provide more precise segmentation. There are several variations of U-Net, some of which obtain state-of-the-art results \cite{cardiacReview}, such as Attention U-Net, which combines a U-shaped architecture with attention gates \cite{oktay2018attention}. Other modifications of vanilla U-Net pretend to improve its segmentation ability or efficiency by replacing its backbone (encoder part).

\section{Convolutional Neural Networks for LVNC measurement}\label{sec:methods}

\subsection{Network architectures}\label{sec:networks}

Given that trabecular measurement using neural networks is an issue that has never been addressed, it is unclear which architecture would achieve the best performance. For that reason, this work proposes the usage of four different network architectures in the experiments, whose results are compared to identify which one performs better for this problem. These four architectures are as follows:

\begin{itemize}
\item The vanilla U-Net \cite{ronneberger_fischer_brox_2015}, as it is widely used for cardiac image segmentation \cite{cardiacReview} and it is the basis of many state-of-the-art solutions for medical segmentation.
\item Attention U-Net \cite{oktay2018attention}. It includes attention gates into the U-Net architecture. Attention gates help the model to focus on the most important areas of the image by reducing the contribution of non-relevant regions.
\item Efficient U-Net B1, which uses the architecture EfficientNet-B1 \cite{tan2019efficientnet} as a substitute for the encoder part of classical U-Net. This network is included as a lightweight architecture alternative to the other models.
\item Pyramid Attention Network (PAN) \cite{li2018pyramid}. It is a lightweight encoder-decoder network, which uses attention mechanisms (similar to the attention gates but more lightweight) and presents an alternative to spatial pyramid pooling. We included this network because this kind of blocks (attention and pyramid pooling) have been used in cardiac image segmentation, achieving state-of-the-art results in some cases \cite{cardiacReview}.
\end{itemize}

In order to estimate and compare models goodness, three kind of evaluations metrics have been utilized: a) Inference time when running both in GPU and CPU, b) Segmentation capabilities (based on Dice coefficient), and c) Diagnosis ability based on network output.

It is important to notice that none of the used architectures takes into account temporal or spatial context.

\subsection{Non-Compaction Measurement and LVNC Diagnosis}

\begin{figure}[h]
    \centering
    \includegraphics[width=54mm]{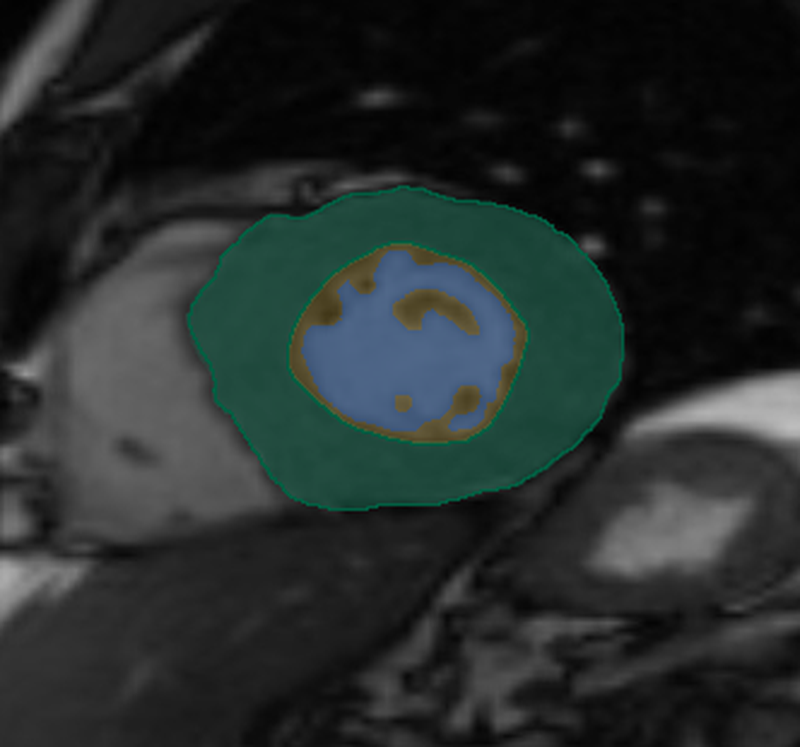}
    \caption{Segmentation sample of left ventricle. Green: external layer. Yellow: trabeculae or non-compacted tissue. Blue: internal cavity}
    \label{fig:desired_segment}
\end{figure}

The principal purpose of this work is to diagnosis LVNC in a fully automatic way, based on available data, i.e. heart images (more specifically MRI slices). In order to measure LVNC levels, three different regions of the left ventricle must be distinguished (as shown in \autoref{fig:desired_segment}): external layer (EL), internal cavity (IC) and trabeculae (T).
 The area of these parts is then used to compute the percentage quantification of the trabecular zone with respect to the compacted area with the following expression \cite{gbernabe}:
\begin{equation}\label{eq:percent_expression}
PTA = 100 \cdot \frac{TA}{TA+ELA} \quad [\%]
\end{equation}
where:
\begin{conditions}
 TA =\text{trabeculae area} \quad \text{and} \quad ELA & external layer area   
\end{conditions}
Although IC information does not appear explicitly in the above expression, we included it as a separate class to distinguish it from the background.

Once the value of expression \eqref{eq:percent_expression} is known for an image, the slice is classified using a threshold: values greater than or equal to 27.4\% are considered as a LVNC indicative \cite{BERNABEGARCIA2017405}.

\subsection{Data}\label{sec:data}

The original data set was composed of 2857 heart short-axis MRI slices of 292 patients, which were stored in DICOM files. All cases had previously been diagnosed with hypertrophic cardiomyopathy (HCM) which is another cardiomyopathy in which myocardium becomes abnormally thick.

Magnetic resonance studies were performed in two hospitals with different scanners:


\begin{itemize}
\item Magnetic Resonance of the Southeast. Affidea (Spain). Imaging was done on a 1.5T scanner (SIGNA HDxt 1,5: General Electric Systems, USA).  Image dimension is $224\times224$ for every subject and pixel spacing of $1.75\times1.75\times0.8mm$ with $2 mm$ between slices.

\item University Hospital Virgen de la Arrixaca in Murcia (Spain). Imaging was done on a 1.5T magnet (Achieva CV, Philips Medical Systems, Netherlands).  Image dimension is $256\times256$ for every subject and pixel spacing of $1.5\times1.5\times0.8mm$ with $2 mm$ between slices.
\end{itemize}

Studies in both hospitals were done without contrast agent, in synchronization with the ECG and patients were holding their breath (at the end of an exhale).  Moreover, slices generated with the first scanner were resampled to a resolution of $256\times256$, as it corresponds to the smallest pixel spacing.

Given the importance of the data set in neural networks training, image slices were successively filtered in a process which we describe in the next lines.

Firstly, we discarded slices where left ventricle was not visible, which resulted in the reduction of the data set to 2438 slices. 

Then, we proceeded to the annotation step. In this research, no traditional and time expensive fully manual annotation was performed. Instead of that, images were labeled using the tool QLVTHC developed in \cite{gbernabe} to segment the whole data set, and cardiologists just manual tuned its parameters for each slice. By doing this, we pretend to overcome a typical issue of this field: this kind of projects usually requires a human expert to segment the whole data set manually. The segmentation produced by this tool has a resolution of $512\times 512$. 

After that, each segmentation produced by QLVTHC, which contained the three regions showed in \autoref{fig:desired_segment}, was evaluated by two cardiologists using the 1 to 5 scale proposed in \cite{gibson_spann_woolley_2004} that is also used in \cite{gbernabe}. This scale allows to measure the quality of the segmentation regarding diagnosis. Thus, a value higher than or equal to $4.0$ indicates that the segmentation does not represent diagnostically significant differences, while values lower than or equal to $3.5$ indicate that it does. Finally, only images with a score higher than or equal to 4.5 were kept.

Once the selection based on cardiologists evaluation was performed, the dataset was reduced to 2362 images. 

The last stage of the filtering process aims to reduce a potential problem derived from resampling: resizing ground-truth segmentation (original size of $512\times 512$) could lead into least sharp class maps. In order to quantify the deformation degree of annotated segmentation, two criteria were applied:

\begin{itemize}
\item For each slice, we computed the percentage of the external layer, the percentage of the trabeculae and the expression \eqref{eq:percent_expression}  before and after the resampling. Images for which the relative error of the resampled segmentation was above $5\%$ for any of these percentages were discarded.

\item As a measure of topological correctness, we computed the number of connected components of the trabecular area before and after the resampling. Only slices for which this value did not differ were kept.
\end{itemize}

These two criteria were met by 2100 MRI slices corresponding to 277 patients, which conformed the final data set used for training. A total of 977 slices corresponded to patients suffering from LVNC.

In addition to the above explained image filtering, another one was performed after the first training. More detailed explanation about it can be found in \autoref{subsec:first-results}.

\subsection{Training}\label{sec:training}

We considered whole-heart models, i.e. separated models for apical, mid-cavity and basal slices were not trained. The resampled MRI slices are normalized to have zero mean and unit variance before they are fed to each network. Moreover, data augmentation was performed for the training set by applying random rotations of 90, 180 and 270 degrees with a $0.25$ probability.

The loss function which was used in all experiments consists of a linear combination of two components:

\begin{equation}\label{eq:loss_expression}
\mathcal{L} = 2\cdot\mathcal{L}_{L} + \mathcal{L}_{B}
\end{equation} 
where
\begin{conditions2}
\mathcal{L}_L & \text{ is the Lovász-Softmax loss defined in \cite{berman_triki_blaschko_2018}.} \\
\mathcal{L}_B & \text{ is a boundary loss inspired in \cite{kervadec2018boundary}. It is computed only for trabeculae.}
\end{conditions2}

With this expression of $\mathcal{L}$, we pretend to obtain a soft loss function to speed up and facilitate learning while focussing on the minimization of Sørensen–Dice coefficient (by using $\mathcal{L}_L$). With $\mathcal{L}_B$  we intend to help addressing the class imbalance problem and to make the output of the networks more anatomically plausible.

Because of the presence of several layers using ReLU activation function in all of the proposed architectures, we use Kaiming initialization to overcome gradient vanishing \cite{he_zhang_ren_sun_2015}. For minimizing the mentioned loss function, Radam optimizer was used, with an initial learning rate of $0.001$ and a weight decay of $0.0005$. Finally, a batch size of 24 was used for every training.

\section{Results and Discussion}\label{sec:results}

The evaluation platform is equipped with a double socket CPU Intel Xeon E5-2603 v3 1.60GHz (12 cores and 64 GB RAM), and two different NVIDIA GPUs: the GeForce RTX 2080 Ti (11GB RAM, 1.350 GHz) for the training, and the GeForce GTX 1080 (8 GB RAM, 1,607 GHz) for the inference.

\subsection{Network Training}

As has been mentioned in \autoref{sec:networks}, four different architectures were evaluated for the problem of LVNC diagnosis. As we can see in \autoref{tab:num_params}, there are great differences concerning the number of parameters of each architecture. It is noteworthy that Efficient U-Net B1 has less than half parameters than the second smallest network.

\sisetup{separate-uncertainty}
\begin{table}[h]
\centering
\ra{1.25}
\begin{tabular}
{@{}
l
S[group-separator={,}, table-format = 8.0, table-text-alignment=right]
S[group-separator={,}, table-format = 2.0, table-text-alignment=right, table-number-alignment = center]
@{}}
\toprule
                   & \multicolumn{1}{c}{Num. Params} & \multicolumn{1}{c}{Training time (min)}  \\ \midrule
U-Net              & 31383876 & 42 \\
Attention U-Net    & 34877616 & 45 \\
Efficient U-Net B1 & 8344804  & 26 \\
PAN                & 21470411 & 28 \\ \bottomrule
\end{tabular}
\caption{Number of parameters of each model and average training time in minutes using a GPU Nvidia GeForce RTX 2080 Ti.}
\label{tab:num_params}
\end{table}

For training, a 5-fold cross-validation process was used, so the data set was split into five equal-sized folds or subsets in a way that all of them had the same proportion of slices of patients with LVNC (stratified cross-validation). The same folds were used for every architecture.
Thus, for each neural network, we obtain five models to evaluate the architecture:  one fold is used as a test set, and the model is trained using the other four-folds (which are splited in train-validation with 80\%-20\%), and this process is repeated for each fold. 
 
Once the model had been trained, it was evaluated over the test set (the left-out fold). Thus, the results discussed below correspond to the whole data set, but the output for each image was produced by a model that had never seen it before.

The training strategy was the same for every fold and architecture: a maximum of 25 epochs is established, but an early stopping can be triggered if loss function $\mathcal{L}$ does not improve during 5 epochs. None of the training processes reached the twenty-fifth epoch, obtaining a maximum of 22 epochs for vanilla U-Net. Due to disparities in the number of parameters, we also found variations in training time for different architectures, as shown in \autoref{tab:num_params}, which contains (in the second column) the average training time of the five training processes for each network.

\subsection{First Results and Data Refinement}
\label{subsec:first-results}

\begin{figure}[h]
\centering
\begin{subfigure}[b]{0.33\textwidth}
\centering
        \includegraphics[width=0.9\linewidth]{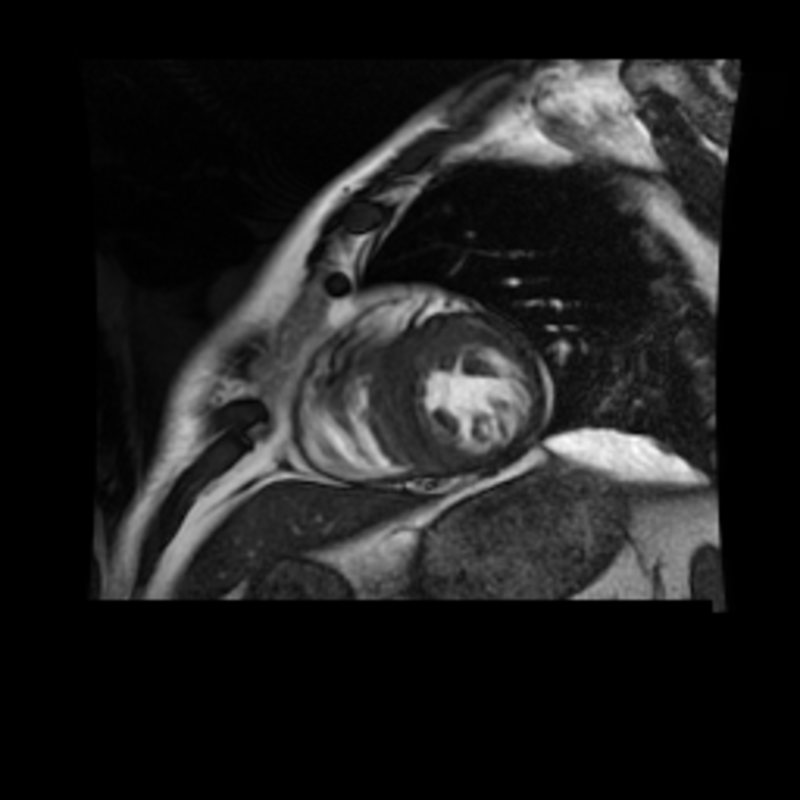}
        \caption{MRI slice}
        \label{fig:gull}
\end{subfigure}\hfill
\begin{subfigure}[b]{0.33\textwidth}
\centering
        \includegraphics[width=0.9\linewidth]{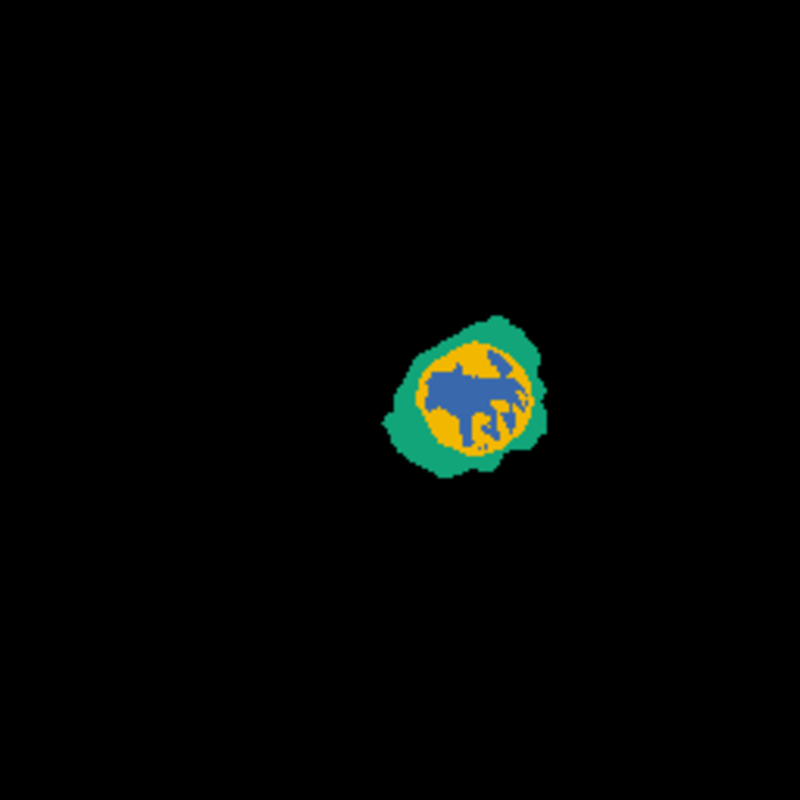}
        \caption{Ground truth}
        \label{fig:gull2}
\end{subfigure}\hfill
\begin{subfigure}[b]{0.33\textwidth}
\centering
        \includegraphics[width=0.9\linewidth]{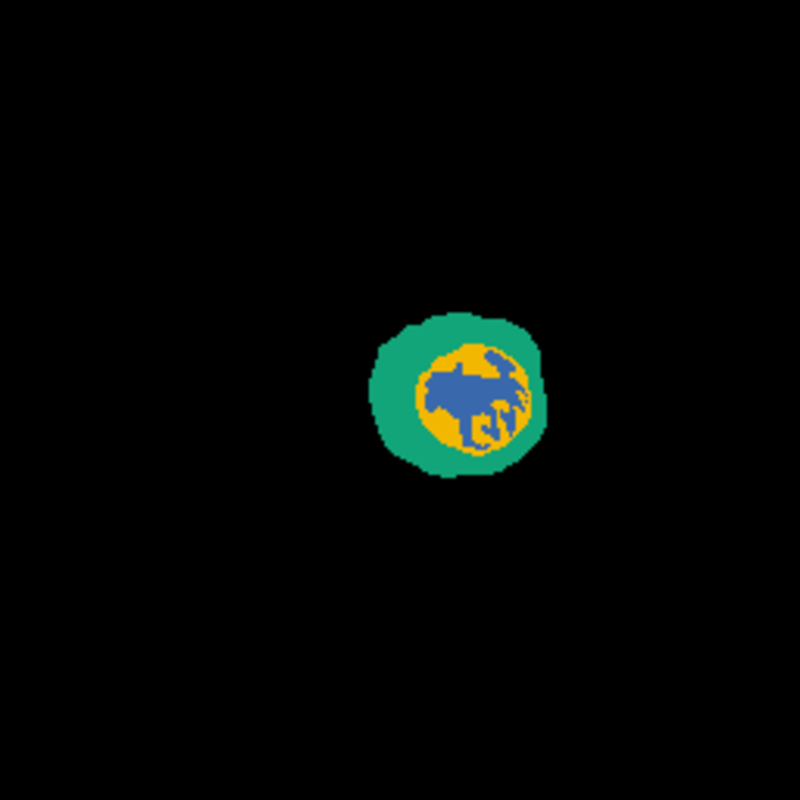}
        \caption{U-Net output}
        \label{fig:tiger}
\end{subfigure}
\caption{Model outputs for a sample slice}\label{fig:bad_ground_truth}
\end{figure}

After the first 5-fold training of every model, outputs were analyzed. It was found that generally, segmentations produced by the trained neural networks were more anatomically plausible than ground truth generated by QLVTHC. Moreover, for specific slices, the output of each of the proposed models was significantly more accurate than reference segmentations (example in \autoref{fig:bad_ground_truth}). While it is true that these deficiencies in ground truth segmentations are not relevant from the diagnosis point of view, the presence poor quality data could potentially add some bias and prevent the networks from learning and generalizing correctly. Therefore, it was hypothesized that discarding images for which the output of trained models was better than ground truth could lead to more accurate models. A total of 80 images were identified and subtracted from the data set, and a new 5-fold training was performed for each architecture. 

\begin{figure}[h]
\centering    
\centering
\includegraphics[width=0.95\textwidth]{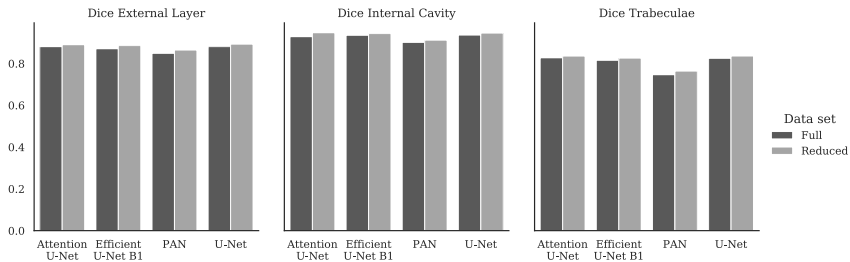}
\caption{Dice coefficient of EL, IC and T regions of left ventricle for every architecture, trained with both the full data set and the reduced data set.}
\label{fig:compare_ignore_1}
\end{figure}

Next, we evaluated the impact of removing these slices. Both models were trained with the full data set (composed of 2100 MRI slices), and the reduced data set (2020 images). The evaluation was carried out only over the reduced one to obtain a fair comparison. \autoref{fig:compare_ignore_1} shows the values of the Dice coefficient for each of the regions considered in the segmentation task. The obtained Dice score is higher when training without the removed slices for every architecture and the three regions. For this reason, the rest of the results and discussion in this section deals with models trained with the reduced data set.

\subsection{Inference Time and Segmentation Results}

Although most of the papers about neural networks mainly focus on the time taken by the training process, we find the measurement inference time equally relevant. The reason for this is that the main purpose of this research is to provide cardiologists with an automated tool that they could use with the available equipment and that helps them by saving time (also when comparing with existing tools). Therefore, the time required for inference was measured when running over CPU and GPU, and it can be found in \autoref{tab:inference_times}: the first column contain results for the Intel CPU, and the second one for the NVIDIA GPU GeForce GTX 1080.

\sisetup{quotient-mode=fraction} 
Both time metrics were obtained from 100 runs with a previous warm-up process of 5 runs. Each one of the architectures performs inference in less than one second when using CPU, but we can highlight both PAN and Efficient U-Net B1 whose inference time is below \SI{0.2}{\second}. Focusing on the GPU results, the benefits are even more evident: thanks to batch processing, a typical MRI study of 15 slices could be completed in less than \SI{0.05}{\second} in any of the networks.

In order to better understand the impact of results in \autoref{tab:inference_times} in time savings for cardiologists, a comparison with existing techniques is performed. Three methods are compared with the proposed models for the already mentioned problem of analyzing 15 slices: fully manual segmentation, QLVTHC semi-automatic tool, and the fully automated tool proposed in \cite{gbernabe}, SOST. Times for fully manual and semi-automatic approaches (25 minutes and 10 seconds per slice, respectively) have been reported by cardiologists. The execution time of SOST has been collected from \cite{gbernabe}, and we picked up the lowest times declared in the paper to achieve a fairer comparison, given that the same machine could not be used. Execution times of both traditional tools correspond to CPU execution, as GPU version of them does not exist.

\begin{figure}[h]
\centering    
\centering
\includegraphics[width=0.95\textwidth]{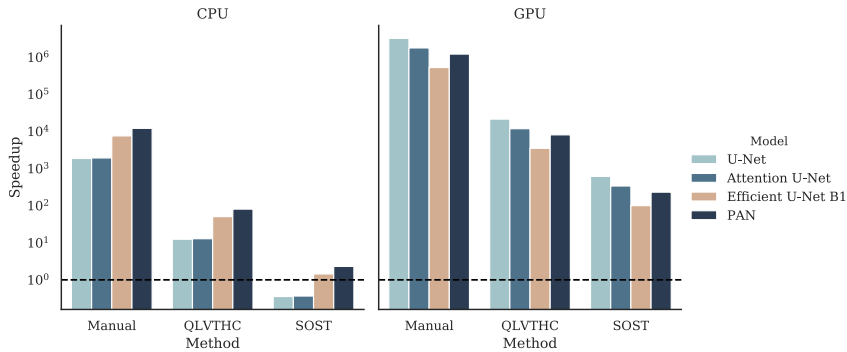}
\caption{Estimated speedups for the problem of quantifying  \textit{PTA} of 15 slices with regard to manual annotation and both manually configured tool (QLVTHC) and fully automated tool (SOST \cite{gbernabe}). Dashed line indicates unit speedup.}
\label{fig:speed_ups}
\end{figure}

As we can see in \autoref{fig:speed_ups}, GPU execution of our models clearly outperforms the three methods, with speedups of up to 
six orders of magnitude with respect to traditional manual segmentation, and up to \SI[round-precision = 0, group-separator={,}, group-minimum-digits=4]{612} when comparing with SOST. In regards with CPU executions, our models also provide great time saving with respect to manual segmentation (up to \SI[round-precision = 0, group-separator={,}, group-minimum-digits=4]{12068} times faster), and the two more lighweight models (PAN and Efficient U-Net B1) are faster than the existing automatic solution SOST. Therefore, these results reinforce the idea that applying accurate and reliable fully automatic tool would lead to a significant time saving for specialists.

\sisetup{separate-uncertainty}
\begin{table}[h]
\centering
\ra{1.25}
\begin{tabular}
{@{}
l
S[table-text-alignment=right, table-format = 3.1(1), round-precision = 2]
S[table-text-alignment=right, table-format = 3.1(1), round-precision = 2]
@{}}
\toprule
                   &  \multicolumn{1}{c}{Time CPU (ms)} & \multicolumn{1}{c}{Time GPU (ms)} \\ \midrule
U-Net              & 804.6 (72) &  7.0 (1) \\
Attention U-Net    &  779.8 (287) & 12.7 (1)  \\
Efficient U-Net B1 &  198.4 (154) & 42.9 (13) \\
PAN                & 124.3 (168) & 18.6 (1) \\ \bottomrule
\end{tabular}
\caption{Mean and standard deviation of inference time for a single slice on CPU, and same metrics for batches of 24 slices on GPU.}
\label{tab:inference_times}
\end{table}

In \autoref{tab:results_ignore}, metrics about the quality of segmentations of every model are displayed. The Dice coefficients of the three U-Net based architectures are very similar, and it is only worthy of note the differences in the error of computed PTA. On the other hand, the difference between the three U-Net based models and PAN is more significant. According to Dice scores, a non-compacted area (trabeculae) seems to be the most difficult to identify by each one of the models. While these differences could be due to class imbalance, given that loss function was chosen to prevent this issue, we are inclined to believe that variations in these values result from the difference in the intrinsic difficulty of recognizing each region.

\sisetup{separate-uncertainty, detect-weight}
\begin{table}[h]
\ra{1.25}
\centering
\begin{tabular}
    {@{}
    l
    S[table-format = 1.2(1), round-precision = 2]
    S[table-format = 1.2(1), round-precision = 2]
    S[table-format = 1.2(1), round-precision = 2]
    S[table-format = 1.2(1), round-precision = 2]
    @{}}
\toprule
                   & \multicolumn{1}{c}{PTA Error (\%)} & \multicolumn{1}{c}{Dice EL} & \multicolumn{1}{c}{Dice IC} & \multicolumn{1}{c}{Dice T}  \\ \midrule
U-Net              &  4.20(469)  &  0.89 (11) &  0.96(13) & 0.84(16) \\
Attention U-Net    & 4.45(554)  & 0.89 (11) & 0.95(10) & 0.84(15) \\
Efficient U-Net B1 & 4.31(517) & 0.89(10) & 0.94(09) &  0.83(15) \\
PAN                & 4.65(563) & 0.86(11) & 0.91(10) & 0.76(15) \\ 
\bottomrule
\end{tabular}
\caption{Mean metrics ($\pm$ standard deviation) of every model. Error committed when computing PTA based on each network output and Dice score for external layer (EL), internal cavity (IC) and trabeculae (T).}
\label{tab:results_ignore}
\end{table}

\subsection{Diagnosis}

Once a slice is segmented, the presence of LVNC is predicted based on the rule of 27.4\% threshold for PTA. \autoref{tab:ignore_metrics} contains some metrics which are usually used to compare diagnosis models for every architecture taken into account. Recall and specificity are included to provide a better understanding of each model's performance regarding positives and negatives separately. On the other hand, accuracy is calculated because it is easy to understand, and it is widely used in literature; however, Matthew coefficient is used as the global comparison metric of the four architectures given that it is more informative than the F1 score and accuracy \cite{Chicco2020}. U-Net outperforms the three others concerning accuracy and Matthews coefficient, and its specificity and recall are slightly lower than the best values. Attention U-Net achieves the best specificity value of the four models, but it also obtains a recall significantly lower than the other three architectures. For its part, Efficient U-Net B1 attains the highest recall and presents the second-best accuracy and Matthews coefficient right behind U-Net. Finally, PAN got the worst results in three of the four metrics, but its recall is very similar to the best one. 

\begin{table}[h]
\ra{1.25}
\centering
\begin{tabular}{
@{}
l
S[table-format = 1.3, round-precision = 3,table-text-alignment=right, table-number-alignment = center]
S[table-format = 1.3, round-precision = 3,table-text-alignment=right, table-number-alignment = center]
S[table-format = 1.3, round-precision = 3,table-text-alignment=right, table-number-alignment = center]
S[table-format = 1.3, round-precision = 3,table-text-alignment=right, table-number-alignment = center]
@{}}
\toprule
                   & \multicolumn{1}{c}{Accuracy} & \multicolumn{1}{c}{\begin{tabular}[c]{@{}c@{}}Matthews\\ Coefficient\end{tabular}} & \multicolumn{1}{c}{Recall} & \multicolumn{1}{c}{Specificity}\\ \midrule
U-Net              & \B 0.865 & \B 0.727 & 0.836 &  0.889 \\
Attention U-Net    & 0.850 & 0.698 & 0.797          & \B 0.894 \\
Efficient U-Net B1 & 0.852           & 0.703                       & \B 0.841 & 0.862 \\
PAN                & 0.847        & 0.691                       & 0.839 & 0.853 \\ \bottomrule
\end{tabular}
\caption{Accuracy, Matthews coefficient, recall and specificity of diagnosis models based on each neural network using the rule of 27.4\% PTA threshold.}
\label{tab:ignore_metrics}
\end{table}

Because of the results showed in \autoref{tab:results_ignore} and \autoref{tab:ignore_metrics}, U-Net presents the best performance because it achieves the best average Dice coefficient for each region and the best Matthews coefficient as a diagnosis model. It is also noteworthy that, despite Efficient U-Net B1 is substantially less complex than the others in terms of the number of parameters, its performance is close to U-Net and Attention U-Net. Moreover, Efficient U-Net B1 inference on CPU is more than four times faster than U-Net, and it is even faster than traditional automated methods, so we consider that this architecture should be taken into account as an efficient alternative to U-Net.

Therefore, outputs of both U-Net and Efficient U-Net B1 were evaluated by two cardiologists in order to determine the validity of the proposed models clinically. The evaluation was performed using the scale previously commented in \autoref{sec:data} \cite{gibson_spann_woolley_2004}. It is important to notice that despite differences in quantitative metrics visible in \autoref{tab:results_ignore} and \autoref{tab:ignore_metrics}, both architectures obtained the same punctuation for each one of the slices. This fact indicates that our selection of the best two models that should be evaluated by specialists is coherent in the sense that their outputs are diagnostically equivalent in spite of being so different regarding the number of parameters.

None of the outputs produced by either U-Net or Efficient U-Net B1 presented diagnostically significant issues. More specifically, $88.84\%$ of the cases were evaluated with the highest score ($5.0$), meaning that cardiologists found the generated segmentation to be completely accurate. The rest of the images obtained an score of $4.5$, so no noticeable differences were found either.

If we take into account results obtained by SOST (reported in \cite{gbernabe}), our two selected models outperform this previously existing automatic tool. Indeed, our full-heart models achieve better cardiologists evaluation even than apical, basal, and mid-cavity specific models of SOST. In particular, the percentage of images whose output obtained a score equal or higher than 4.5 was 78.22\% for the best SOST model (only valid for mid-cavity slices) and 63.16\% for the full-heart SOST model; in comparison with the 100\% of the two networks selected in this paper. It is noteworthy that images utilized in the evaluation of SOST in \cite{gbernabe} constitute a subset of the data set used during this research.

\section{Conclusions}\label{sec:conclusion}

In this paper, we have proposed four different convolutional neural networks as a first approach to LVNC measurement using deep learning techniques. The three U-Net alternatives performed better than PAN architecture for segmentation and diagnosis metrics. We could remark two architectures: firstly, vanilla U-Net, which obtained the best results in almost every metric. On the other hand, Efficient U-Net B1, which presented similar metrics to U-Net, and its number of parameters, training time and inference time on CPU are significantly lower than other architectures.

U-Net and Efficient U-Net B1 were evaluated by cardiologists taking into account the goodness of their outputs for diagnosis, i.e. a subjective evaluation of the accuracy of the generated segmentations. Each one of the slices was found to be perfectly accurate for diagnosis, and 88.84\% of them was evaluated as a perfect match.

Moreover, using the two finally selected models, segmentations can be obtained in less than \SI{0.2}{\second} when using a CPU and in less than \SI{0.01}{\second} on a GPU. These times corroborate that this kind of tools could lead to considerable time savings of the specialists, given that fully manual segmentation takes them about 25 minutes per slice.

Results achieved confirm the main idea that motivated this research: deep learning based approaches can achieve excellent results in the diagnosis and measurement of LVNC. In fact, proposed vanilla U-Net and Efficient U-Net B1 solutions achieve improvements in both execution time and quality of outputs (based on cardiologists evaluation) when compared with already existing fully automatic method.

In spite of these promising results, this paper presents only a first approach to this problem, and critical challenges remain. Firstly, the trabecular area presents the lowest Dice coefficient in every considered case. Given that the problem we are trying to solve is the quantification of trabeculae, exporing some options to improve this metric should be a priority. Moreover, it could be interesting to use techniques like weakly supervised learning to work with poor quality data, like the ground truth segmentations generated by already existing tools.

\begin{acknowledgements}
This work was supported by the Spanish MCIU and AEI, as well as
European Commission FEDER funds, under grant RTI2018-098156-B-C53.
\end{acknowledgements}

%
%

\bibliographystyle{spbasic}      
\bibliography{references}   

\end{document}